\documentclass[12pt,aps,prb,showpacs,amsmath,amssymb]{revtex4}

\usepackage{graphicx} 
\usepackage{dcolumn} 
\usepackage{amsmath}
\usepackage{amssymb}

\def\sg{{\sigma}}

\def\vk{{\bf k}}

\def\bra{\langle}
\def\ket{\rangle}

\def\eps{\epsilon}

\newcommand{\eq}[1]{Eq.~(\ref{#1})}
\newcommand{\fig}[1]{Fig.~\ref{#1}}
\newcommand{\be}{\begin{equation}}
\newcommand{\ee}{\end{equation}}
\newcommand{\bea}{\begin{eqnarray}}

\newcommand{\eea}{\end{eqnarray}}
\newcommand{\bean}{\begin{eqnarray*}}
\newcommand{\eean}{\end{eqnarray*}}
\newcommand{\bfi}{\begin{figure}}
\newcommand{\efi}{\end{figure}}
\newcommand{\bc}{\begin{center}}
\newcommand{\ec}{\end{center}}
\newcommand{\ba}{\begin{array}}
\newcommand{\ea}{\end{array}}

\begin{document}

\title{Mean-field theory for symmetry-breaking \\ 
 Fermi surface deformations on a square lattice}

\author{Hiroyuki Yamase$^{1}$, Vadim Oganesyan$^{2}$, and 
 Walter Metzner$^{1}$} 
\affiliation{
 $^{1}$Max-Planck-Institute for Solid State Research, Heisenbergstrasse 1, 
 D-70569 Stuttgart, Germany, \\
 $^{2}$Department of Physics, Princeton University, Princeton, 
 NJ 08544, USA}

\date{\today}

\begin{abstract}

We analyze a mean-field model of electrons with pure forward scattering 
interactions on a square lattice which exhibits spontaneous Fermi
surface symmetry breaking with a $d$-wave order parameter:
the surface expands along the $k_x$-axis and shrinks along the
$k_y$-axis (or vice versa).
The symmetry-broken phase is stabilized below a dome-shaped
transition line $T_c(\mu) \,$, with a maximal $T_c$ near van Hove
filling.
The phase transition is usually first order at the edges of the 
transition line, and always second order around its center.
The $d$-wave compressibility of the Fermi surface is however
strongly enhanced even near the first order transition down to 
zero temperature.
In the weak coupling limit the phase diagram is fully determined 
by a single non-universal energy scale, and hence dimensionless 
ratios of different characteristic quantities are universal.
Adding a uniform repulsion to the forward scattering 
interaction, the two tricritical points at the ends of the second 
order transition line are shifted to lower temperatures.
For a particularly favorable choice of hopping and interaction 
parameters one of the first order edges is replaced completely
by a second order transition line, leading to a quantum critical 
point.
\end{abstract}

\pacs{71.18.+y, 71.10.Fd}

\maketitle


\section{Introduction}
 
The low energy properties of an interacting electron system are 
strongly influenced by the shape of its Fermi surface.
Residual interactions between quasi-particles near the Fermi
surface can give rise to charge- or spin-density waves, 
superconductivity, or other low energy instabilities.
Usually the Fermi surface respects the point-group symmetry of
the underlying lattice structure. 
In principle, however, electron-electron interactions can 
drive a Fermi surface deformation that breaks the orientational 
symmetry of the system.

Recently, the possibility of symmetry-breaking Fermi surface
deformations with a $d$-wave order parameter, where the surface 
expands along the $k_x$-axis and shrinks along the $k_y$-axis 
(or vice versa), was discussed for various two-dimensional 
electron models on a square lattice: $t$-$J$,\cite{yamase00} 
Hubbard,\cite{halboth00,grote02,neumayr03} and extended
Hubbard\cite{valenzuela01} model.
The instability is driven by interactions in the forward 
scattering channel, mainly between electrons close to the van
Hove points in the two-dimensional Brillouin zone.
Referring to Pomeranchuk's\cite{pomeranchuk58} stability 
condition for isotropic Fermi liquids it has been termed 
\emph{''Pomeranchuk instability''} by some authors.
Only a discrete lattice symmetry is broken by the $d$-wave
deformation on the square lattice, 
in constrast to spontaneous Fermi surface symmetry breaking 
in isotropic Fermi liquids, where Goldstone modes play an 
important role.\cite{oganesyan01}
The $d$-wave Fermi surface deformation leads to a state with 
the same reduced symmetry as the \emph{''nematic''} electron liquid
defined by Kivelson {\it et al.}\cite{kivelson98} in a discussion 
of possible analogies between doped Mott insulators with charge 
stripe correlations and liquid crystal phases.\cite{kivelson03}

Spontaneous Fermi surface symmetry breaking competes with other 
instabilities.
In the slave-boson mean-field theory of the $t$-$J$ model the $d$-wave
Fermi surface deformation is overwhelmed by $d$-wave singlet
pairing, but strongly enhanced correlations in the $d$-wave forward 
scattering channel and a corresponding large response to external 
anisotropic perturbations remain.\cite{yamase00,yamase04} 
The latter can be related to distinctive properties of magnetic 
excitations in different cuprate superconductors.\cite{yamase00,
yamase01}
In the Hubbard model near van Hove filling coexistence of 
superconductivity and $d$-wave Fermi surface symmetry breaking 
has been found in a renormalized weak coupling perturbation 
expansion.\cite{neumayr03}

Quantum critical fluctuations of the ''soft'' Fermi surface near
a continuous zero temperature phase transition with Fermi surface 
symmetry breaking provide a route to non-Fermi liquid 
behavior.\cite{metzner03}
Anomalously large and anisotropic quasi-particle decay rates have 
been derived for a phenomenological model, where electrons moving 
on a square lattice interact only via almost forward scattering 
interactions, that is only very small momentum transfers are 
allowed.\cite{metzner03}
We refer to this model as the \emph{''f-model''} in the following.
Recently it was shown that the putative quantum critical
point in the f-model is actually preempted by a first order 
transition, at least within mean-field theory and for various
concrete choices of the model parameters.\cite{kee03}
The transition remains first order at low finite temperatures
but turns to second order at temperatures above a tricriticial
point.\cite{khavkine04}

This paper is dedicated to a comprehensive mean-field analysis 
of spontaneous Fermi surface symmetry breaking in the f-model 
on a square lattice. 
We present results for the phase diagram, order parameter and 
Fermi surface as obtained from a numerical
solution of the mean-field equations for various typical choices
of parameters. The weak coupling limit is analyzed analytically.
We also compute the coefficients of the Landau expansion of the
grand canonical potential up to quartic order in the order 
parameter and show that the first order transition at low 
temperature is a rather robust consequence of the van Hove 
singularity in the density of states.
Besides confirming the conclusions by Kee {\it et al.}\cite{kee03} 
and Khavkine {\it et al.}\cite{khavkine04} and providing additional
numerical data especially at finite temperatures, we present
several new results and insights.
In particular, we show that in the weak coupling limit Fermi
surface symmetry breaking is characterized by a single energy scale,
which leads to universal behavior in terms of suitably rescaled 
parameters.
Furthermore, we show that the tricritical points can be suppressed 
to lower temperatures by a uniform repulsion added to the original 
f-model, which, for a particularly favorable but not unphysical 
choice of hopping and interaction parameters, can even lead to a 
quantum critical point.
Finally, we find that the $d$-wave compressibility of the Fermi surface 
is usually strongly enhanced along the transition line down to zero 
temperature even if the transition is first order, which implies that 
the Fermi surface is already very soft at the transition and 
fluctuations should be important.

The article is structured as follows.
In Sec.\ II we define the f-model and outline the mean-field theory
of spontaneous Fermi surface symmetry breaking. 
Results from a numerical solution of the mean-field equations are
presented in Sec.\ III. 
The numerical results are complemented by an analysis of the
Landau free energy expansion in Sec.\ IV and an analytic derivation 
of universal properties of the phase transition at weak coupling 
in Sec.\ V.
We finally conclude in Sec.\ VI.



\section{Model and mean-field theory}

We analyze the f-model on a square lattice with pure forward 
scattering interactions. The Hamiltonian reads
\begin{equation}
 H = \sum_{\vk} \epsilon_{\vk}^{0} \, n_{\vk} + 
 \frac{1}{2L} \sum_{\vk,\vk'} f_{\vk\vk'} \, n_{\vk} n_{\vk'}
 \label{f-model}
\end{equation}
in standard second quantized notation, where 
$n_{\vk} = \sum_{\sg} n_{\vk\sg}$ counts the spin-summed number
of electrons with momentum $\vk$, and $L$ is the number of lattice
sites.
For hopping amplitudes $t$, $t'$, and $t''$ between nearest,
next-nearest, and third-nearest neighbors on the square
lattice, respectively, the bare dispersion relation is given by
\begin{equation}
 \epsilon_{\vk}^{0}= -2 \left[t (\cos k_{x}+\cos k_{y}) 
 + 2t'\cos k_{x} \cos k_{y} +t''(\cos 2k_{x} + \cos 2k_{y}) \right]
 \; . 
\end{equation}
The forward scattering interaction has the form 
\begin{equation}
 f_{\vk\vk'} = u - g \, d_{\vk} d_{\vk'} \,,   \label{fkk}
\end{equation}
with coupling constants $u \geq 0$ and $g>0$, and a function 
$d_{\vk}$ with $d_{x^2-y^2}$-wave symmetry such as 
$d_{\vk} = \cos k_x - \cos k_y$.
This ansatz mimics the structure of the effective interaction in 
the forward scattering channel as obtained for the 
$t$-$J$,\cite{yamase00} Hubbard,\cite{halboth00} and extended
Hubbard\cite{valenzuela01} model.
The uniform term originates directly from the repulsion between 
electrons and suppresses the (uniform) electronic compressibility 
of the system. 
The $d$-wave term enhances the $d$-wave compressibility of the
Fermi surface and drives spontaneous Fermi surface symmetry 
breaking.
In the Hubbard model it is generated by (1-loop) fluctuations,
while in the $t$-$J$ and extended Hubbard model the nearest 
neighbor interaction contributes directly to a $d$-wave 
attraction in the forward scattering channel.
For $u=0$, the above model is the pure forward scattering limit
of the f-model with small momentum transfers introduced in 
Ref.\ \onlinecite{metzner03}.
Within mean-field theory, it is also equivalent to the model
analyzed in Refs.\ \onlinecite{kee03,khavkine04}, since the 
off-diagonal components of the quadrupole density introduced 
there do not affect the results.

Inserting $n_{\vk} = \bra n_{\vk} \ket + \delta n_{\vk}$ into
the interacting part of the model and neglecting terms of order 
$(\delta n_{\vk})^2$, we obtain the mean-field Hamiltonian
\begin{equation}
 H_{\rm MF} = \sum_{\vk} \epsilon_{\vk} \, n_{\vk}
 - \frac{1}{2} \sum_{\vk} \delta\epsilon_{\vk} \,
 \bra n_{\vk} \ket \,,
 \end{equation}
where 
$\epsilon_{\vk} = \epsilon^0_{\vk} + \delta\epsilon_{\vk}$ 
is a renormalized dispersion relation, which is shifted with 
respect to the bare dispersion by
\begin{equation}
 \delta\epsilon_{\vk} = 
 \frac{1}{L} \sum_{\vk'} f_{\vk\vk'} \, \bra n_{\vk'} \ket =
 u \, n + \eta \, d_{\vk} \; .
\end{equation}
Here $n = L^{-1} \sum_{\vk} \bra n_{\vk} \ket$ is the average
particle density, and 
\begin{equation}
 \eta = 
 - \frac{g}{L} \sum_{\vk} d_{\vk} \bra n_{\vk} \ket
\end{equation}
is our order parameter, which parametrizes the amount of symmetry
breaking. 
Note that $\eta$ is real and has the dimension of energy.
It vanishes as long as the momentum distribution function 
$\bra n_{\vk} \ket$ respects the symmetry of the square lattice.
The grand canonical potential per lattice site 
$\omega = L^{-1} \Omega$
is obtained from the mean-field Hamiltonian as
\begin{equation}
 \omega = \frac{\eta^{2}}{2g} - \frac{u}{2} n^{2} -
 \frac{2T}{L} \sum_{\vk} \log(1+e^{-(\epsilon_{\vk} - \mu)/T}) 
 \; .
 \label{freeenergy}
\end{equation}
The stationarity conditions 
$\frac{\partial\omega}{\partial\eta} = 0$ and
$\frac{\partial\omega}{\partial n} = 0 \,$ (at fixed $\mu$) 
yield the self-consistency equation for the order parameter 
\begin{equation}
 \eta = - \frac{2g}{L} \sum_{\vk} 
 d_{\vk} f(\epsilon_{\vk}-\mu)
 \label{selfeta}
\end{equation}
and the equation determining the density
\begin{equation}
 n = \frac{2}{L} \sum_{\vk} f(\epsilon_{\vk}-\mu) \; ,
 \label{selfn} 
\end{equation}
respectively, where $f(\xi) = (e^{\xi/T} + 1)^{-1}$ is the Fermi 
function. These equations follow also directly from the relation
$\bra n_{\vk} \ket = 2\, f(\epsilon_{\vk}-\mu)$.
Note that the Eqs.\ (\ref{selfeta}) and (\ref{selfn}) are coupled
for $u \neq 0$, since $n$ enters the self-consistency equation 
for $\eta$ via $\epsilon_{\vk}$.

In the thermodynamic limit the mean-field theory solves the reduced
version (no momentum transfers) of the f-model exactly.
One simple way to see this is by considering the Feynman diagrams 
representing the perturbation expansion of the system. 
All self-energy diagrams except the Hartree term involve integrals 
over momentum transfers and are thus suppressed at least by a 
factor $L^{-1}$.


\section{Numerical Results}

We first take band parameters $t'/t = -1/6$ and $t'' = 0$, and 
solve the self-consistency equations eqs.~(\ref{selfeta}) and 
(\ref{selfn}) numerically. 
For this choice of hopping amplitudes the bare dispersion relation
has saddle points at $(\pi,0)$ and $(0,\pi)$, leading to a van
Hove singularity in the bare density of states at 
$\epsilon^0_{\rm vH} = 4t' = -2t/3$. 
Typical features of spontaneous Fermi surface symmetry breaking are 
captured with these parameters and are presented in the first three 
subsections. 
In the last subsection we investigate another set of hopping
parameters, for which the saddle points are slightly shifted from 
$(\pi,\,0)$ and $(0,\,\pi)$, and a quantum critical point can be
realized for suitable choices of the interaction parameters.

\subsection{Typical phase diagram}

\begin{figure}[t]
\centerline{\includegraphics[scale=0.69]{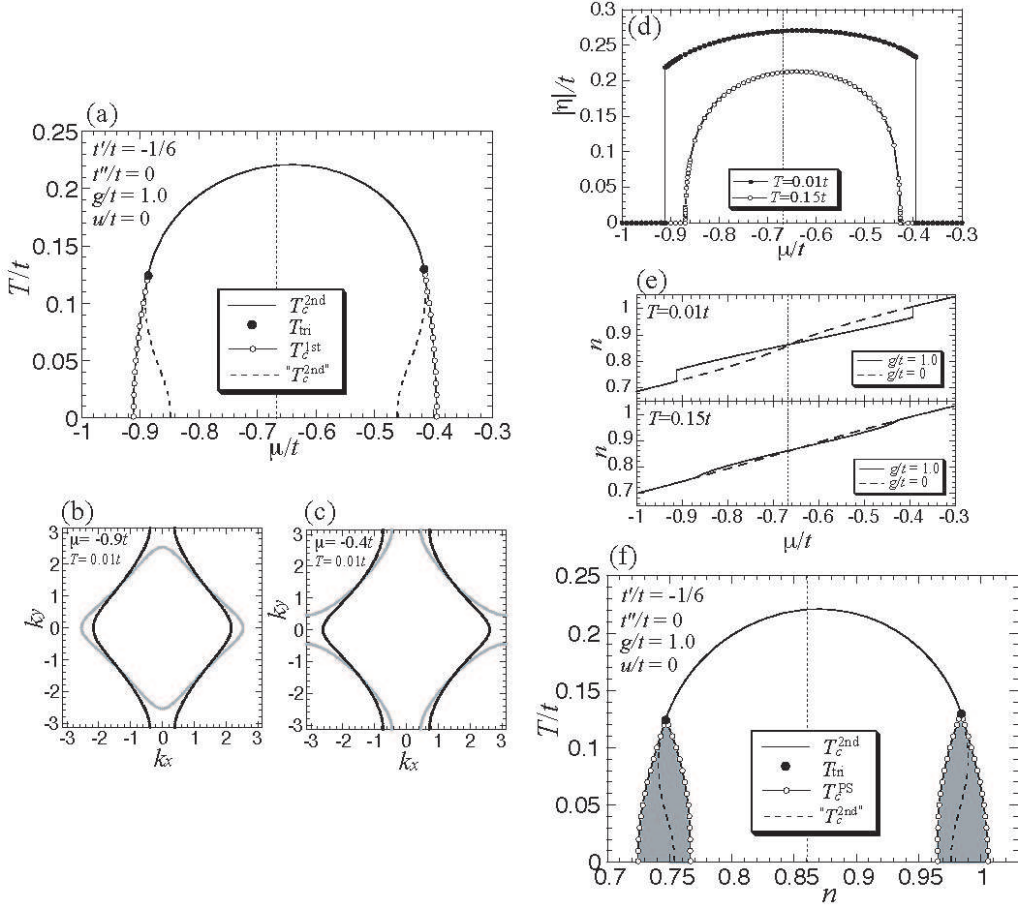}}
\caption{The mean-field solution for $t'/t=-1/6$, $t''/t=0$, 
$g/t=1$, and $u=0$. 
(a) $\mu$-$T$ phase diagram;
the total transition line, $T_{c}(\mu)$, contains a second order 
transition line, $T_{c}^{\rm 2nd}$, at high temperatures and two 
first order lines, $T_{c}^{\rm 1st}$, at low temperatures; 
the solid circles are tricritical points; the dashed line, 
''$T_{c}^{\rm 2nd}$'', denotes a fictitious second order transition 
that is preempted by the first order transition; the dotted line 
indicates the van Hove energy, $\mu = \epsilon_{\rm vH}^0 = -2t/3$. 
(b) and (c) Fermi surface in the symmetry-broken phase near the 
first order transition; the Fermi surface for $g=0$ is also shown 
by a gray line.  
(d) $\mu$ dependence of $|\eta|$ at $T=0.01t$ and $0.15t$. 
(e) $\mu$ dependence of $n$ at $T=0.01t$ and $0.15t$; 
the results for $g=0$ are plotted also. 
(f) $n$-$T$ phase diagram; $T_{c}^{\rm 2nd}$ is a second order 
transition temperature and solid circles are tricritical points. 
In the shaded regions, which are surrounded by $T_{c}^{\rm PS}$, 
the system undergoes phase separation.} 
\label{phase}
\end{figure}

We first focus on the $d$-wave term in \eq{fkk} and set $u=0$.  
A $\mu$-$T$ phase diagram is shown in \fig{phase}(a). 
The solid line denotes a second order phase transition, which 
turns to a first order transition at low $T$ (open circles).
The end points of the second order transition are tricritical 
points (solid circles), where the quadratic and quartic 
coefficients of the Landau energy expansion (see Sec.\ IV)
vanish simultaneously. 
The dashed line denotes the fictitious second order transition  
that is preempted by the first order transition. 
The complete transition line, $T_{c}(\mu)$, has a dome shape. 
Since the Fermi surface symmetry breaking is driven by forward 
scattering of electrons mainly on the (original) Fermi surface 
close to the van Hove points,\cite{yamase00,halboth00}   
the maximal $T_{c}(\mu)$ appears for $\mu$ around the van Hove 
energy $\epsilon^{0}_{\rm vH}=-2t/3$; 
a slight deviation from $\epsilon^{0}_{\rm vH}$ is due to 
finite $T$ effects. 
The transition line is almost symmetric with respect to the 
$\mu=\epsilon_{\rm vH}^{0}$ axis. 
This symmetry becomes exact when $t'$ is set to zero because of
particle-hole symmetry.  
Below $T_c$ the Fermi surface expands along the $k_{x}$-direction 
and shrinks along the $k_{y}$-direction or vice versa. We show 
results for the Fermi surface at low $T$ in Figs.~\ref{phase}(b) 
and (c) together with the corresponding bare Fermi surface ($g=0$) 
for comparison. The Fermi surface has typically open topology 
in the symmetry-broken phase, except close to the second order 
transition. 

In \fig{phase}(d), we plot the order parameter $|\eta|$ as a 
function of $\mu$ at $T=0.01t$ and $0.15t$. 
While we see a continuous transition at $T=0.15t$, 
$|\eta|$ exhibits a jump at $T=0.01t$, characteristic of a 
first order transition. 
In \fig{phase}(e), the density $n$ is plotted as a function of 
$\mu$ at $T=0.01t$ and $0.15t$ together with that for $g=0$. 
The density increases monotonuously with $\mu$. This behavior 
is due to the stability condition of the system that the grand 
canonical potential must be a concave function as a function 
of $\mu$,  which yields an inequality, 
$-\frac{\partial^{2}\omega}{\partial^{2}\mu} >0$, 
or $\frac{\partial n}{\partial \mu} >0$. 
The density changes discontinuously at the first order transitions. 
The directions of the density jumps are generic features required 
by the concavity of the grand canonical potential. 

In \fig{phase}(f) the phase diagram is plotted in the $n$-$T$ plane.
The second order transition line at high $T$ terminates at two 
tricritical points, below which phase separated regions, shown 
by shades, appear; 
dashed lines correspond to the ficticious second order transition
shown already in \fig{phase}(a). 
Each shaded region  has two phases, with different densities 
$n_{1}$ and $n_{2} (>n_{1})$. 
The difference between $n_{1}$ and $n_{2}$ corresponds to the 
magnitude of the density jump at the first order transition 
point [see \fig{phase}(e)].
For $n_{1}\leq n \leq n_{2}$ the volume fraction of the low 
density phase and the high density phase is 
$\frac{n_{2}-n}{n_{2}-n_{1}}$ and $\frac{n-n_{1}}{n_{2}-n_{1}}$,
respectively. 
The former has a symmetric (symmetry-broken) Fermi surface and 
the latter a symmetry-broken (symmetric) Fermi surface in the 
left (right) phase separated region.

\subsection{Effects of uniform repulsion} 

\begin{figure}[t]
\centerline{\includegraphics[scale=0.7]{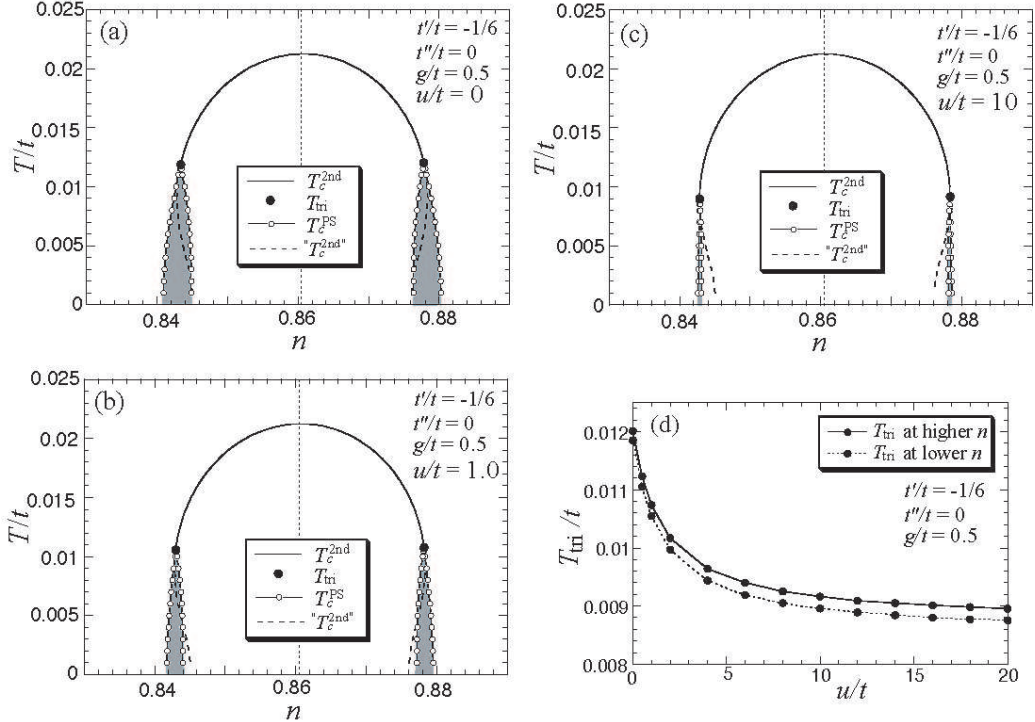}}
\caption{$n$-$T$ phase diagram for 
$t'/t=-1/6$, $t''/t=0$, and $g/t=0.5$ with (a) $u=0$, (b) $u/t=1.0$, and 
(c) $u/t=10$. (d) $u$-dependence of $T_{\rm tri}$ at higher 
$n$ (solid line) and at lower $n$ (dotted line).}
\label{uphase}
\end{figure}

Now we switch on the uniform term in the interaction, \eq{fkk}. 
The $n$-$T$ phase diagrams are shown in Figs.~\ref{uphase}(a)-(c) 
for several choices of $u$ at $g=0.5t$. 
Note that typical features of the phase diagrams are the same as 
in \fig{phase} although we take smaller $g$ and finite $u$ here. 
The second order transition line is not affected by $u$ in the 
$n$-$T$ phase diagram, while it would be affected if plotted in
the $\mu$-$T$ plane, because the chemical potential corresponding
to a given density varies with $u$. 
The tricritical points, that is the end points of the second order 
transition, extend to lower $T$ with $u$ in favor of a second order 
transition, which is accompanied by a pronounced suppression of 
the width of the phase separated regions, since $n_2 - n_1$ is 
strongly reduced by the uniform term in the interaction. 
However, we see that the suppression of $T_{\rm tri}$ saturates at 
large $u$, as shown in \fig{uphase}(d), and is not strong enough 
to establish a quantum critical point.

\subsection{{$\boldsymbol d$}-wave compressibility near the first order 
            transition} 

When the symmetry-broken phase is realized through a first order 
transition at low $T$, order parameter fluctuations are not 
critical at the transition.
However, we now show that the anisotropic compressibility with
a $d$-wave form factor is strongly enhanced by interactions
at the transition line, such that fluctuations can be expected 
to be important in spite of the first order character of the 
transition.

The $d$-wave compressibility 
\begin{equation}
 \kappa_d = \frac{d n_d}{d \mu_d}
 \label{dcomp}
\end{equation}
describes the linear response of the expectation value
$n_d = L^{-1} \sum_{\vk} d_{\vk} \, \bra n_{\vk} \ket$ to the
symmetry-breaking perturbation
$H_d = - \mu_d \sum_{\vk} d_{\vk} \, n_{\vk}$.
The perturbation $H_d$ induces a $d$-wave shaped deformation of
the Fermi surface.
Note that the order parameter $\eta$ is directly proportional 
to $n_d$, namely $\eta = - g n_d$.
Without the form factors $d_{\vk}$ the above expressions would
yield the conventional compressibility 
$\kappa = \frac{dn}{d\mu}$.
For our mean field model, the $d$-wave compressibility is given 
exactly by the RPA expression 
\begin{equation}
 \kappa_d =\frac{N_2}{1 - gN_2} \, , \label{rpa}
\end{equation}
where $N_2$ is a weighted density of states with $d_{\vk}^2$
as a weight factor [see \eq{wdos} in Sec.\ IV].
The denominator
\begin{equation}
 S^{-1} = 1 - gN_2 
\end{equation}
is the inverse ''Stoner factor'', which is a dimensionless
measure for the enhancement of the $d$-wave compressibility
by interactions, and hence for the enhancement of order
parameter fluctuations.
 
\begin{figure}[t]
\centerline{\includegraphics[scale=0.4]{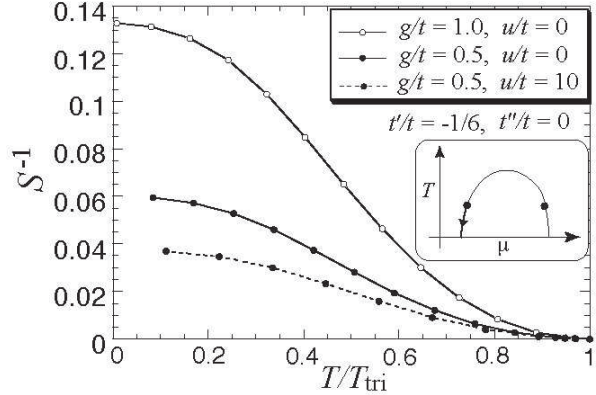}}
\caption{The inverse of the Stoner enhancement factor
for several choices of $g$ and $u$ along a first order 
transition line as sketched in the inset with the arrow;  
temperature is scaled by the tricritical temperature.}
\label{stonerfig}
\end{figure}

We calculate $S^{-1}$ along one of the two first order transition 
lines as sketched in the inset of \fig{stonerfig};
similar results are obtained along the other side of the first 
order transition. 
The main panel of \fig{stonerfig} shows that $S^{-1}$ tends to
zero at the tricritical temperature, that is the compressibility 
$\kappa_d$ diverges as expected, indicating truely critical 
fluctuations. 
At lower temperatures, $S^{-1}$ is finite on the transition line. 
However, its value is still much smaller than one, especially for 
a smaller $g$; the introduction of $u$ reinforces this tendency. 
The $d$-wave compressibility is thus strongly enhanced by 
interactions at the first order transition line down to the
lowest temperatures, for example by a factor of about 25 for 
$g=0.5t$ and $u=10t$. 
Hence, Fermi surface and thus order parameter fluctuations can 
be expected to be important even near the first order transition 
at low $T$. 
In the weak coupling analysis presented in Sec.\ V we will show 
that $S^{-1}$ can be arbitrarily small for small $g$ near the first 
order transition. 

\subsection{Quantum critical point} 

\begin{figure}[t]
\centerline{\includegraphics[scale=0.8]{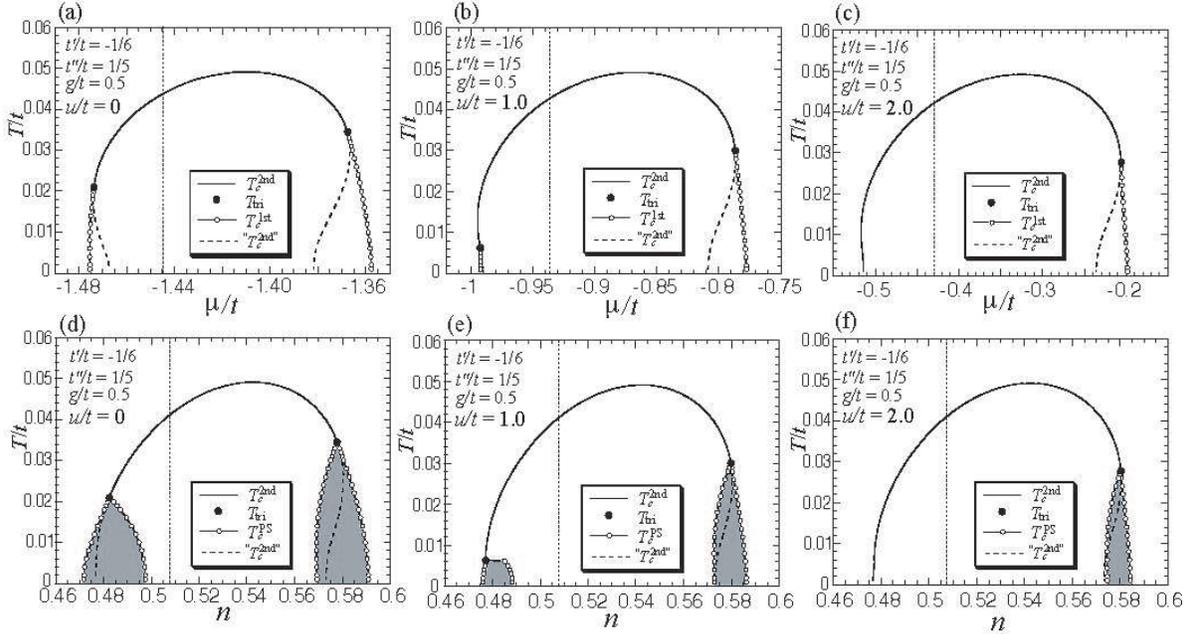}}
\caption{$\mu$-$T$ phase diagrams for several choices of $u$ for 
$t'/t=-1/6$, $t''/t=1/5$, and $g/t=0.5$: $u/t=0$ (a), $1.0$ (b), and 
$2.0$ (c). The dotted lines indicate the van Hove energy. The lines 
of $T_{c}^{\rm 1st}$ and ``$T_{c}^{\rm 2nd}$'' at lower $\mu$ in (b) 
are too close to be distinguished. 
The  $\mu$-$T$ phase diagrams are replotted as a function of $n$ 
in (d)-(f).}
\label{Yphase}
\end{figure}

The above results for our choice of hopping parameters, 
$t'/t = -1/6$ and $t''/t = 0$, are probably rather generic in the
sense that other choices lead to qualitatively the same phase
diagrams, as long as the bare kinetic energy has saddle points 
at  $(\pi,0)$ and $(0,\pi)$.
Another interesting set of parameters is $t'/t = -1/6$ and 
$t''/t = 1/5$. 
For $t'' > \frac{1}{4} (t + 2t')$ the bare kinetic energy has
saddle points at $(\pi\pm\cos^{-1}\alpha,\,0)$ and 
$(0,\,\pi\pm\cos^{-1}\alpha)$ with $\alpha=(t+2t')/4t''$, and  
local minima at $(\pi,0)$ and $(0,\pi)$. 
The energy at the saddle points is given by  
$\epsilon^{0}_{\rm vH} = -2(t-2\alpha^{2}t'')$.
For $t'/t=-1/6$ and $t''/t=1/5$ one has $\alpha = \frac{5}{6}$
and $\epsilon^{0}_{\rm vH} = - \frac{13}{9} t$.

A sequence of phase diagrams for $t'/t=-1/6$ and $t''/t=1/5$ is 
plotted in \fig{Yphase}; the coupling $g$ is $0.5t$ in all cases.
The transition line $T_c(\mu)$ for $u=0$ in \fig{Yphase}(a) has a 
dome shape similar to that in \fig{phase}(a), with a second order 
transition at high $T$ and a first order transition at low $T$. 
Although the Fermi surface symmetry breaking is mainly due to 
electrons close to the Fermi surface near the van Hove points, 
the maximum of $T_{c}(\mu)$ largely deviates from the van Hove 
energy and the dome shows a pronounced asymmetry. 
This is related to a large asymmetry of the density of states 
(see Sec.\ IV). 
A striking feature of the present parameters is a drastic 
suppression of one of the tricritical points by a moderate 
uniform repulsion in the interaction, which leads to a 
quantum critical point for sufficiently large $u$. 
We show $\mu$-$T$ phase diagrams for $u/t=1$ and $2$ in 
Figs.~\ref{Yphase}(b) and (c), respectively.  
The tricritical point at higher $\mu$ is only slightly suppressed 
by $u$ and the suppression saturates at large $u$ as in the case 
of \fig{uphase}. 
However, the first order line at lower $\mu$ is suppressed rapidly 
with increasing $u$ and disappears completely in favor of a
continuous phase transition down to the lowest temperatures beyond 
$u/t \sim 2$, that is a quantum critical point is realized.
Why the choice of hopping parameters with saddle points deviating
from $(\pi,0)$ and $(0,\pi)$ is favorable for a continuous phase
transition at low temperatures will be discussed in connection
with the Landau expansion in Sec.~IV.

In Figs.\ \ref{Yphase}~(d)-(f), we replot the $\mu$-$T$ phase 
diagrams as a function of $n$. The first order transitions at low 
$T$ lead to phase separated regions as already seen in \fig{phase}. 
However, on the lower density side of the $n$-$T$ phase diagram in
\fig{Yphase}(e) the phase separated region opens discontinuously 
at the critical temperature $T_{\rm tri}$, where the second order
line terminates. 
In that case the end point of the second order line is not a 
tricritical point (and $T_{\rm tri}$ thus actually a misnomer).
The quartic term of the Landau energy is positive there, but a
first order transition nevertheless sets in due to a local
minimum at finite $\eta$ in the Landau energy becoming a global
one below $T_{\rm tri}$.

\begin{figure}[t]
\centerline{\includegraphics[scale=0.5]{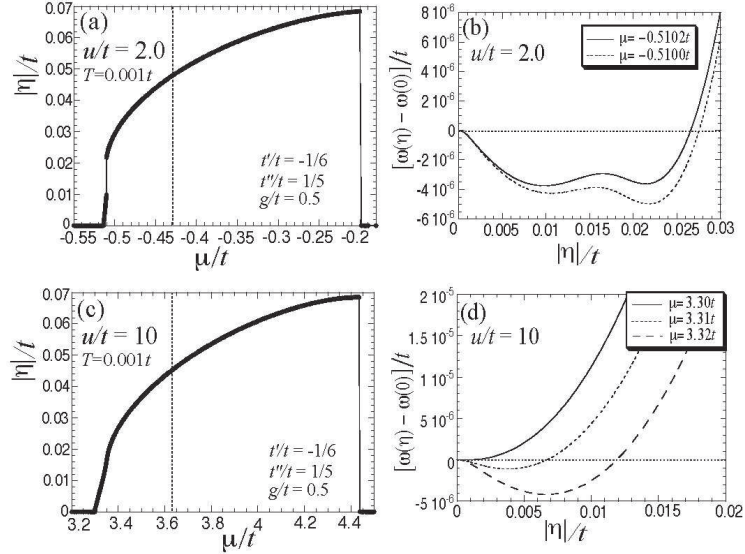}}
\caption{Left panels: Order parameter $|\eta|$ as a function
of $\mu$ for hopping parameters and $g$ as in Fig.~\ref{Yphase},
temperature $T/t = 0.001$, and two different uniform couplings:
$u/t = 2$ (top) and $u/t = 10$ (bottom).
Right panels: Landau energy $\omega(\eta)$ near the phase 
transitions at the lower $\mu$ side of the corresponding
order parameter plots on the left.}
\label{Ynm-eta}
\end{figure}

In Fig.~\ref{Ynm-eta} we show the $\mu$-dependence of the order
parameter $\eta$ for hopping parameters and coupling $g$ as in 
Fig.~\ref{Yphase}, at a very low temperature ($T/t = 0.001$) and
two different choices for $u$. 
This plot reveals that for $u/t = 2$ there is a first order
transition within the symmetry-broken phase in addition and very 
close to the quantum critical point shown in Fig.~\ref{Yphase} 
(c). The corresponding Landau energy [Fig.~\ref{Ynm-eta} (b)]
has two minima at finite $|\eta|$; the minimum at lower $|\eta|$ 
has the lowest energy only for $\mu$ between the second and first 
order transition point.
For larger $u$ the first order transition in the symmetry-broken
phase moves further away from the second order transition until
it disappears completely. 
We show in Fig.~\ref{Ynm-eta} (c) that the order parameter for 
$u/t = 10$ is continuous everywhere except at the first order 
transition at the large $\mu$ boundary of the symmetry-broken
phase. The Landau energy for $\mu$ close to the second order
transition has only one minimum as a function of $|\eta|$ in 
this case [Fig.~\ref{Ynm-eta} (d)].

\begin{figure}[t]
\centerline{\includegraphics[scale=0.33]{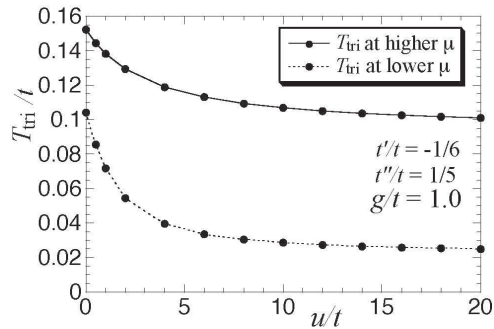}}
\caption{$u$-dependence of $T_{\rm tri}$ at higher $\mu$ (solid line) 
and at lower $\mu$ (dotted line) for 
$t'/t=-1/6$, $t''/t=1/5$, and $g/t=1.0$.}
\label{YTtri}
\end{figure}

No quantum critical point appears for $t'/t=-1/6$ and $t''/t=1/5$ 
if the coupling constant $g$ is too small or too large. 
In \fig{YTtri} we plot $T_{\rm tri}$ as a function of $u$ for 
$g/t=1.0$. Although $T_{\rm tri}$ at lower $\mu$ is suppressed 
much stronger than that at higher $\mu$, the suppression saturates 
at a finite temperature even for large $u$. 
Remarkably the Stoner factor $S$ is again generally strongly 
enhanced for the present hopping parameters near the transition 
at higher $\mu$, while $S$ is not large at low $T$ 
near the transition at lower $\mu$ for $u=0$, although 
$T_{\rm tri}$ is lower there. Obviously a low $T_{\rm tri}$
does not imply a large Stoner factor.


\section{Landau expansion}

\begin{figure}[t]
\centerline{\includegraphics[scale=0.45]{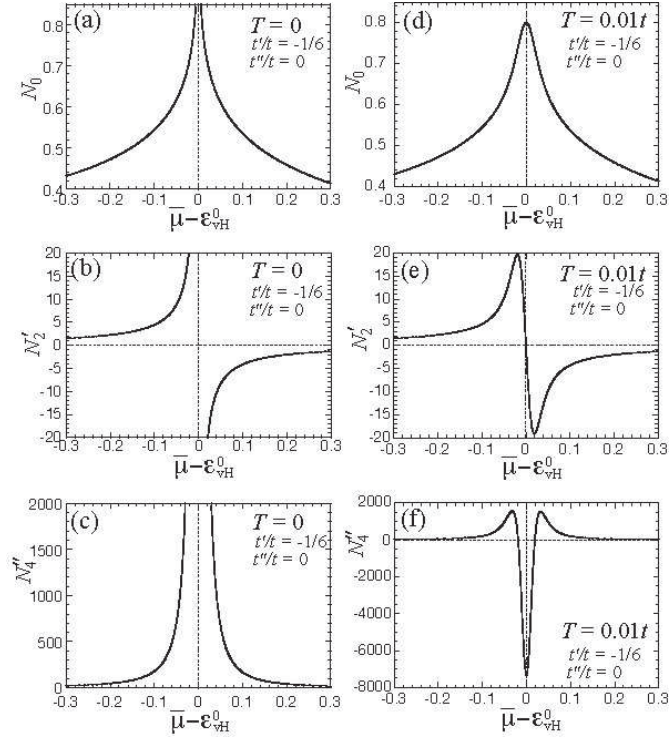}}
\caption{The functions $N_{0}(\bar\mu,T)$, $N_{2}'(\bar\mu,T)$, 
and $N_{4}''(\bar\mu,T)$ around the van Hove energy   
at $T=0$ [(a)-(c)] and $T=0.01t$ [(d)-(f)] for hopping
parameters $t'/t = -1/6$ and $t''=0$; 
the energy unit is $t$.}
\label{nmu}
\end{figure}

To gain a broader understanding of the symmetry-breaking phase 
transitions and their dependence on model parameters we consider 
the Landau expansion of the grand canonical potential in powers 
of the order parameter
\begin{equation}
 \omega(\eta) - \omega(0) = 
 \frac{1}{2} \, a_{2} \, \eta^{2} + 
 \frac{1}{4!} \, a_{4} \, \eta^{4} + \dots \label{GL}
\end{equation}
up to quartic order. The function $\omega(\eta)$ is given by
Eq.\  (\ref{freeenergy}), with the density $n(\eta)$ determined 
by Eq.\ (\ref{selfn}). 
Note that only even powers appear due to the symmetry 
$\omega(-\eta) = \omega(\eta)$. 
Expanding $\omega(\eta)$ by taking $\eta$-derivatives (see
Appendix) one obtains the coefficients
\begin{eqnarray}
 a_{2} &=& g^{-1} - N_2(\bar\mu,T) \; , \label{a2} \\
 a_{4} &=& - N_{4}''(\bar\mu,T) + 
 \frac{3u}{1 + u \, N_{0}(\bar\mu,T)}[N_{2}'(\bar\mu,T)]^{2} \; , 
 \label{a4}
\end{eqnarray}
where $\bar\mu = \mu - u \, n(0)$ and
\begin{equation} \label{wdos}
 N_p(\bar\mu,T) = - \, \frac{2}{L} \sum_{\vk} d_{\vk}^p \,
 f'(\epsilon_{\vk}^0 - \bar\mu)
\end{equation}
is a weighted density of states averaged over an energy interval 
of order $T$ around $\bar\mu$, and $N'_p$, $N''_p$ are first 
and second derivatives with respect to $\bar\mu$.
Note that $a_2$ depends only via $\bar\mu$ on $u$, which explains
the $u$-independence of the second order transition lines in the
$(n,T)$ phase diagrams in Figs.~\ref{uphase} and \ref{Yphase}.
The quartic coefficient $a_4$ does not depend on $g$.
In Fig.~\ref{nmu} we plot $N_{0}(\bar\mu,T)$, $N_{2}'(\bar\mu,T)$, 
and $N_{4}''(\bar\mu,T)$ for $T=0$ and $T=0.01t$, for the choice
of hoppings underlying the results in Sec.\ IIIA-C. 
At zero temperature $N_p(\bar\mu,T)$ diverges logarithmically for
$\bar\mu \to \epsilon^0_{\rm vH}$, which leads to 
$N_{2}'(\bar\mu,T) \propto - (\bar\mu - \epsilon^0_{\rm vH})^{-1}$ and 
$N_{4}''(\bar\mu,T) \propto (\bar\mu-\epsilon^0_{\rm vH})^{-2}$ near 
$\epsilon^0_{\rm vH}$. These singularities are cut off for 
$|\bar\mu-\epsilon^0_{\rm vH}| < T$ at finite temperature.

The critical manifold in the space spanned by $g$, $\bar\mu$, and 
$T$, on which a continuous phase transition may occur, is given 
by the condition $a_2 = 0$, that is $g \, N_2(\bar\mu,T) = 1$. 
However, the continuous transition can be realized only if the
quartic coefficient $a_4$ is positive. Otherwise it will be 
preempted by a first order transition.
For $u=0$, one has simply $a_4 = - N''_4(\bar\mu,T)$, which is 
obviously negative near the van Hove singularity at temperatures 
$T \ll |\bar\mu-\epsilon^0_{\rm vH}|$, and positive at any $T>0$ for
$\bar\mu = \epsilon^0_{\rm vH}$. 
Hence, the transition is first order at temperatures 
$T \ll |\bar\mu-\epsilon^0_{\rm vH}|$, but it can be expected to
be continuous near the maximum of $T_c$ around van Hove filling, 
as is indeed the case in all numerical results.
For $u>0$ there is an additional positive contribution to $a_4$,
which partially compensates the negative main term. 
However, the positive term is bounded by 
$3[N_{2}'(\bar\mu,T)]^{2}/N_0(\bar\mu,T)$ even for arbitrarily 
large $u$.
For $\bar\mu$ near $\epsilon^0_{\rm vH}$ it is of order
$1/\log|\bar\mu-\epsilon^0_{\rm vH}|$ smaller than 
$N''_4(\bar\mu,T)$.

At moderate distance from van Hove filling a numerical evaluation
reveals that for hopping parameters $t'/t = -1/6$ and $t''=0$
the positive $u$-term can compensate a substantial amount of the 
negative term, $- N''_4(\bar\mu,T)$, but $a_4$ never turns positive.
We thus confirm that a finite $u$ shifts the tricritical points
in the phase diagrams to lower temperatures, but does not produce 
a continuous phase transition at $T=0$ for that choice of hopping.

\begin{figure}[t]
\centerline{\includegraphics[scale=0.45]{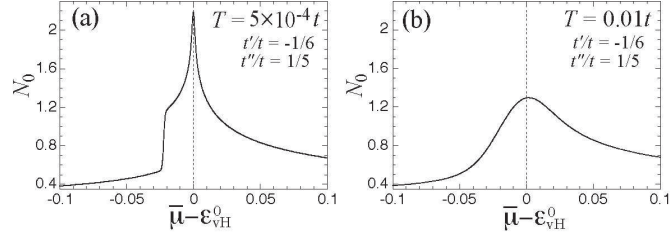}}
\caption{$N_{0}(\bar\mu,T)$ around the van Hove energy   
at $T=5\times 10^{-4}t$ (a) and $T=0.01t$ (b) for hopping
parameters $t'/t = -1/6$ and $t''=1/5$; 
the energy unit is $t$.}
\label{Ynmu}
\end{figure}

The situation is very different for our second choice,
$t'/t = -1/6$ and $t''/t = 1/5$. 
In that case $N_p(\bar\mu,T)$ has a step-like increase at the lower
$\mu$ side of the van Hove energy at low $T$, which is generated by 
the local minima in the dispersion at $(\pi,0)$ and $(0,\pi)$.
This is illustrated for $N_0$ in \fig{Ynmu}; $N_2$ and $N_4$ behave
similarly.
For $\bar\mu$ in the step region $N'_2(\bar\mu,T)$ becomes very 
large at low $T$, while $N_0(\bar\mu,T)$ remains bounded. 
In the presence of a $u$-term it is thus possible to get a 
positive $a_4$ at arbitrarily low temperatures. 
A sizable $u$ helps not only to increase $a_4$,
but also to push local minima at finite $\eta$ in $\omega(\eta)$ 
to higher energies, such that a continuous transition can be 
obtained before such a minimum becomes global.

\section{Weak coupling limit}

At weak coupling spontaneous Fermi surface symmetry breaking 
occurs only for densities near van Hove filling, and the transition 
is dominated by states with momenta near the saddle points of 
$\epsilon^0_{\vk}$.
In this limit the mean-field equations can be treated to a certain
extent analytically, and the phase transition is universal in the
sense that it is fully characterized by a single energy scale. 
Several ratios of physical quantities are universal dimensionless
numbers in the weak coupling limit.

We focus on the case $u=0$ and assume that $\epsilon^0_{\vk}$ has 
only two degenerate saddle points in $\vk_A = (\pi,0)$ and 
$\vk_B = (0,\pi)$. However, the following analysis can be easily
extended to energy bands with saddle points in other positions. 
Shifting energies such that $\epsilon^0_{\vk_{A,B}} = 0$ and 
choosing suitable relative momentum variables $k_+$ and  $k_-$,  
one can write $\epsilon^0_{\vk}$ near the saddle points as a 
quadratic form
\begin{equation} \label{quadeps}
 \epsilon^0_{\vk} = \frac{1}{2m} \, k_+ k_- \, ,
\end{equation}
where $m > 0$ is a constant which can be related to the
hopping amplitudes $t$, $t'$, and $t''$.  
The variables $k_+$ and $k_-$ are defined such that the quadratic 
form has no $k_+^2$ and $k_-^2$ terms and that the integration 
measure is $(2\pi)^{-2}$ as usual.
Note that the corresponding coordinate axes are generally not
orthogonal. 
The momenta are restricted to a finite region around the saddle 
points by a cutoff $\Lambda$, that is $|k_{\pm}| \leq \Lambda$. 

If the form factor $d_{\vk}$ is smooth near the van Hove point
it can be taken as a constant with alternating sign near $\vk_A$ 
and $\vk_B$, respectively, such as
\begin{equation}
 d_{\vk} = \left\{ \begin{array}{rll}
 1  & {\rm for} & \vk \approx \vk_A \, ,\\
 -1 & {\rm for} & \vk \approx \vk_B\, .
 \end{array} \right.
\end{equation}
Any other constant ($\neq 1$) could be absorbed in the coupling 
$g$. The self-consistency condition \eq{selfeta} becomes simply
\begin{equation}
 \eta = g \, (n_B - n_A)\,,
\end{equation}
where 
\begin{equation}
 n_{A,B} = 2 \int_{A,B} \frac{d^2k}{(2\pi)^2} \, 
 f(\epsilon^0_{\vk} \pm\eta - \mu)
\end{equation}
is the contribution of the momentum space region around $\vk_A$ 
and $\vk_B$ (limited by $\Lambda$) to the density.

\subsection{Ground state}

At $T=0$ the integral over the Fermi function is simply the 
volume of occupied states. 
An elementary integration yields
\begin{equation}
 n_{A,B} = \frac{\Lambda^2}{\pi^2} + 
 \frac{2m}{\pi^2} \, (\mu \mp \eta) \,
 \left( 1 + \log\frac{\Lambda^2}{2m} - 
 \log|\mu \mp \eta| \right) \, .
\end{equation}
The self-consistency equation can thus be written as
\begin{equation}
 \eta = \bar g \, \big[
 (\mu-\eta) \log|\mu-\eta| -
             (\mu+\eta) \log|\mu+\eta| 
 + 2\eta\left( 1+\log\epsilon_{\Lambda} \right) \big]
\end{equation}
with the dimensionless coupling $\bar g = \frac{2m}{\pi^2} \, g$
and the cutoff energy $\epsilon_{\Lambda} = \Lambda^2/2m$.
Integrating 
$\partial\omega/\partial\eta = g^{-1} \eta - (n_B - n_A)$
over $\eta$ one obtains the $\eta$-dependence of the grand 
canonical potential 
\begin{eqnarray}
 \omega(\eta) &=& \frac{2m}{\pi^2} \left\{ 
 \left[ \frac{1}{2 \bar g} 
 - \frac{1}{2} - (1+\log\epsilon_{\Lambda}) \right] \eta^2 
 \right. \nonumber \\[2mm] && + \left.
 \frac{1}{2} \, (\mu+\eta)^2 \log|\mu+\eta| + 
 \frac{1}{2} \, (\mu-\eta)^2 \log|\mu-\eta| \right\}
 + {\rm const.} \; ,
\end{eqnarray}
where the constant does not depend on $\eta$ and can be chosen
such that $\omega(0) = 0$.

For $\mu = 0$, where the bare Fermi surface touches the van Hove 
points, the self-consistency equation for $\eta$ becomes simply
\begin{equation}
 \eta = - 2 \bar g \, \eta \log|\eta| + 2 \bar g \, \eta \, 
 \log(e \, \epsilon_{\Lambda}) \; .
\end{equation}
Besides the trivial solution $\eta = 0$ this equation has the
two degenerate solutions
\begin{equation}
 \eta_0 = \pm \, e \, \epsilon_{\Lambda} \, e^{-1/(2 \bar g)} \; . 
 \label{eta-g}
\end{equation}
It is easy to see that the solution $\eta = 0$ corresponds to a
maximum of $\omega(\eta)$ and is thus unstable.
The total density remains unaffected by the symmetry breaking, 
since $\delta n_B = - \delta n_A$ for $\mu=0$.
This is also true to high accuracy for the density at van
Hove filling for the numerical result shown in \fig{phase}(e). 
Note that at van Hove filling symmetry breaking occurs for 
arbitrarily small $g$, which is due to the logarithmic divergence
of the density of states.

For $\mu \neq 0$ we introduce a rescaled order parameter 
$\tilde\eta = \eta/\mu$. The self-consistency condition for
$\eta$ can be written in terms of $\tilde\eta$ as  
\begin{equation}
 \frac{\tilde\eta}{\tilde g} = 
 (1-\tilde\eta) \log|1-\tilde\eta| -
 (1+\tilde\eta) \log|1+\tilde\eta| 
\end{equation}
with a renormalized coupling constant given by
\begin{equation}
 \frac{1}{\tilde g} = \frac{1}{\bar g} + 2\log|\mu|
 - 2(1+\log\epsilon_{\Lambda}) \; .
 \label{tildeg}
\end{equation}
The grand canonical potential can be written as
$\omega(\eta) = \frac{2m}{\pi^2} \, \mu^2 \, 
 \tilde\omega(\tilde\eta)$, where
\begin{equation}
 \tilde\omega(\tilde\eta) = 
 \left( \frac{1}{2 \tilde g} - \frac{1}{2} \right) 
 \tilde\eta^2 +
 \frac{1}{2} \, (1+\tilde\eta)^2 \log|1+\tilde\eta| + 
 \frac{1}{2} \, (1-\tilde\eta)^2 \log|1-\tilde\eta| \; .
\end{equation}
Note that $\mu$ and the cutoff have been completely absorbed
in the renormalized coupling $\tilde g$. 
The rescaled self-consistency equation and 
$\tilde\omega(\tilde\eta)$ are universal in the sense that they 
depend only via $\tilde g$ on all input parameters.

Minimizing $\tilde\omega(\tilde\eta)$ one finds that a first 
order transition occurs at the universal critical coupling 
$\tilde g_1 \approx -0.692$ with a universal jump of the 
dimensionless order parameter 
\begin{equation}
 |\tilde\eta_1| = \frac{|\eta_1|}{|\mu_1|} \approx 1.720
\end{equation}
The Fermi surface thus opens immediately at the transition.
Inverting the relation between $\bar g$ and $\tilde g$ one 
obtains the $\mu$-dependence of $\bar g_1$ as
\begin{equation}
 \bar g_1(\mu) = \frac{\tilde g_1}{1 + 2 \tilde g_1 
 (\log\frac{\epsilon_{\Lambda}}{|\mu|} + 1)} 
 \quad \stackrel{\mu \to 0}{\to} \quad 
 \frac{1}{2\log\frac{\epsilon_{\Lambda}}{|\mu|}} \; .
\end{equation}
Note that $\bar g_1 > 0$ for $|\mu| \ll \epsilon_{\Lambda}$ 
although $\tilde g_1 < 0$.
For fixed $g$, on the other hand, the critical value of 
$\mu$ at which the first order transition occurs is 
\begin{equation}
 |\mu_1| = e^{1 + 1/(2\tilde g_1)} \, \epsilon_{\Lambda}
 \, e^{-1/(2 \bar g)} \; .
\end{equation}
The curvature of $\tilde\omega(\tilde\eta)$ at $\tilde\eta = 0$
becomes negative only for $\tilde g > \tilde g_2 = - 0.5$.
At fixed $\mu$ this requires couplings $g > g_2(\mu) > g_1(\mu)$.
For given $g$, the critical value of $\mu$ for a continuous 
transition is 
\begin{equation}
 |\mu_2| = \epsilon_{\Lambda} \, e^{-1/(2 \bar g)} \; ,
\end{equation}
which is smaller than $|\mu_1|$.
Hence, the critical point for a continuous transition is not
reached, but preempted by a first order transition.
The ratio
$|\mu_1|/|\mu_2| = e^{1 + 1/(2\tilde g_1)} \approx 1.320$
is a universal number.

In the symmetric state, the interaction induced enhancement of 
the $d$-wave compressibility of the Fermi surface is given by the 
''Stoner factor'' $S = [1 - g N_2(\mu)]^{-1}$, which is related 
to the quadratic coefficient of the Landau expansion by
$S = (g a_2)^{-1}$, see \eq{a2}. Since $a_2$ vanishes for 
$g = g_2$, one has $a_2 = g^{-1} - g_2^{-1}$. At the first order 
transition in the ground state, $a_2$ is given by
\begin{equation}
 \left. a_2 \right|_1 = g_1^{-1} - g_2^{-1} = 
 \frac{2m}{\pi^2} \, \left(
 {\tilde g_1}^{-1} - {\tilde g_2}^{-1} \right) 
 \approx 0.555 \, \frac{2m}{\pi^2} \, ,
\end{equation}
where we have used \eq{tildeg} in the second step.
The $d$-wave compressibility is thus enhanced by a factor
\begin{equation}
 S_1 = \frac{1}{0.555 \, \bar g} \label{stonereq}
\end{equation}
at the first order transition. 
For a weak coupling $\bar g \ll 1$, this enhancement is very large.

\subsection{Finite temperature}

We now compute two characteric temperature scales in the weak
coupling limit, namely the transition temperature at van Hove
filling, $T_0$, and the tricritical temperature, $T_{\rm tri}$.
To this end we write the functions $N_p(\mu,T)$ in the form
\begin{equation} \label{np1}
 N_p(\mu,T) = - \int d\eps N_p(\eps) \, f'(\eps-\mu) \; ,
\end{equation}
where $N_p(\eps) = N_p(\eps,0)$, which, 
for the quadratic dispersion in \eq{quadeps} with a cutoff
$\Lambda$, is given by
\begin{equation} \label{np2}
 N_p(\eps) = 
 \frac{4m}{\pi^2} \, \log\frac{\eps_{\Lambda}}{|\eps|}
\end{equation}
with $|\eps| \leq \eps_{\Lambda}$.
Note that $\bar\mu = \mu$ for $u=0$.

The transition temperature at van Hove filling is obtained by 
setting the quadratic coefficient in the Landau expansion 
$a_2$ to zero at $\mu = 0$, that is by solving the equation
$g N_2(0,T_0) = 1$.
Using
\begin{equation}
 \int_{-\eps_{\Lambda}}^{\eps_{\Lambda}} d\eps \,
 \log\frac{|\eps|}{\eps_{\Lambda}} \, f'(\eps) \; 
 \stackrel{\eps_{\Lambda}/T \to \infty}{\longrightarrow} \;
 \log\frac{\eps_{\Lambda}}{T} - \log\frac{\pi}{2} + \gamma \; ,
\end{equation}
where $\gamma \approx 0.577$ is the Euler constant, one obtains
\begin{equation}
 T_0 = \frac{2 e^{\gamma}}{\pi} \, \eps_{\Lambda} \,
 e^{-1/(2\bar g)} \; .
\end{equation}
Note that the numerical prefactor coincides precisely with the
one in the BCS formula for the critical temperature of a weak 
coupling superconductor.

At the tricritical point $a_2$ and $a_4$ both vanish.
The tricritical temperature $T_{\rm tri}$ and the corresponding
chemical potential $\mu_{\rm tri}$ are thus determined by the
two equations $g N_2(\mu_{\rm tri},T_{\rm tri}) = 1$ and
$N''_4(\mu_{\rm tri},T_{\rm tri}) = 0$.
Using Eqs.\ (\ref{np1}) and (\ref{np2}) one obtains
\begin{equation} \label{n2asy}
 g N_2(\mu,T) \;
 \stackrel{\eps_{\Lambda}/T \to \infty}{\longrightarrow} \;
 2 \bar g \big[ \log\frac{\eps_{\Lambda}}{T} + a(\tilde\mu) \big] 
 \; ,
\end{equation}
where $\tilde\mu = \mu/T$ and the dimensionless function 
$a(\tilde\mu)$ is defined as
\begin{equation}
 a(\tilde\mu) = 
 \int_{-\infty}^{\infty} dx \, \log|x+\tilde\mu| \, 
 \frac{\partial}{\partial x} \, \frac{1}{e^x + 1} \; .
\end{equation}
Furthermore
\begin{equation}
 N''_4(\mu,T) \; 
 \stackrel{\eps_{\Lambda}/T \to \infty}{\longrightarrow} \;
 \frac{4m}{(\pi T)^2} \, b(\tilde\mu)
\end{equation}
with
\begin{equation}
 b(\tilde\mu) = 
 \int_{-\infty}^{\infty} dx \, \log|x+\tilde\mu| \, 
 \frac{\partial^3}{\partial x^3} \, \frac{1}{e^x + 1} \; .
\end{equation}
The latter function vanishes for 
$\tilde\mu = \pm\tilde\mu_{\rm tri}$ with
$\tilde\mu_{\rm tri} \approx 1.911$.
Setting the right hand side of \eq{n2asy} equal to one and solving
for $T$ then yields
\begin{equation}
 T_{\rm tri} = 
 e^{-\alpha} \, \eps_{\Lambda} \, e^{-1/(2 \bar g)}
\end{equation}
where $\alpha = a(\tilde\mu_{\rm tri}) \approx 0.4515$.
Hence, the tricritical temperature and the critical temperature
at van Hove filling form the universal ratio
\begin{equation}
 \frac{T_{\rm tri}}{T_0} = 
 \frac{\pi e^{-\alpha}}{2 e^{\gamma}} \approx 0.5614 \; .
\end{equation}

\subsection{Comparison with numerical results}

Above we have computed several physical quantities characterizing
the phase transition, which are all proportional to the same
energy scale $\eps_{\Lambda} \, e^{-1/(2 \bar g)}$, with 
universal prefactors in the weak coupling limit. 
Hence ratios of these quantities are universal numbers.

We have checked universal ratios against results from 
the numerical solution of the mean-field equations for hopping 
parameters $t'/t = -1/6$, $t'' = 0$ and coupling $u = 0$.
At zero temperature we have checked ratios involving the order 
parameter $\eta_0$ at van Hove filling, the order parameter jump 
$\eta_1$ at the first order transition, and the distances of 
$\mu_1$ and $\mu_2$ from $\eps^0_{\rm vH}$.
For a comparison of the order parameter one has to take into 
account a factor two due to the different size of $d_{\vk}$
near the van Hove points, which is $\pm 2$ in the numerical
calculation with $d_{\vk} = \cos k_x - \cos k_y$, but $\pm 1$
in the weak coupling model.
At finite temperature we have compared the transition 
temperature $T_0$ at van Hove filling and the tricritical point.
For $g/t = 0.5$ all ratios agree within one percent error with 
the predicted universal numbers. 
For the stronger coupling $g/t = 1$ the deviation increases to 
two or three percent at zero temperature and up to around five
percent at finite temperature.


\section{Conclusion}

In summary, we have analyzed a mean-field model for Fermi surface
symmetry breaking with a $d$-wave order parameter on a square 
lattice.
We have confirmed the qualitative properties of the phase diagram
reported already by Kee {\it et al.}\cite{kee03} and Khavkine 
{\it et al.}\cite{khavkine04} We have provided further numerical 
evidence and analytic arguments showing that the phase transition 
is typically first order at low temperatures.
This implies that a stability analysis of microscopic models,
for example by renormalization group methods, should not focus 
on diverging susceptibilities only.
At weak coupling the transition is fully characterized by a single 
energy scale and can thus be described by universal dimensionless 
functions of suitably rescaled parameters, which leads to various
universal ratios of different quantities.
The tricritical points separating first and second order behavior
are shifted to lower temperatures by adding a repulsive
constant contribution to the forward scattering interaction,
which, for a favorable choice of hopping and interaction 
parameters, can even lead to a quantum critical point.
Although the phase transition is usually first order at low $T$, 
we have found that the $d$-wave compressibility at the transition 
can be enhanced significantly by interactions, which implies that 
Fermi surface fluctuations induced by interactions with a small
finite momentum transfer are expected to be important even near 
the first order transition.
The role of fluctuations, in particular their influence on the 
phase transition, remains an interesting subject for future 
studies.


\begin{acknowledgments}
We are grateful to L.\ Dell'Anna for valuable discussions.
V.O.\ thanks I.\ Khavkine, C.\ Chung and H.\ Kee
for collaboration on a related work, and NSF-DMR-0213706 and the
David and Lucile Packard Foundation grants for support.
H.Y.\ has been partially supported by a special postdoctoral 
researchers program from RIKEN, Japan.
\end{acknowledgments}


\appendix

\section{Derivation of Landau expansion}

Here we derive the expressions Eqs.\ (\ref{a2}) and (\ref{a4}) for
the coefficients of the Landau expansion of $\omega(\eta)$, by 
taking derivatives with respect to $\eta$ at fixed $\mu$ and with
$n(\eta)$ determined by \eq{selfn}.
The first derivative is
\begin{equation}
 \frac{d\omega}{d\eta} = 
 \frac{\partial\omega}{\partial\eta} + 
 \frac{\partial\omega}{\partial n} \frac{d n}{d\eta} =
 \frac{\eta}{g} + 
 \frac{2}{L} \sum_{\vk} d_{\vk} \, f(\epsilon_{\vk} - \mu)
\end{equation}
where we have used the stationarity condition 
$\frac{\partial\omega}{\partial n} = 0$.
The second derivative is
\begin{equation}
 \frac{d^2\omega}{d\eta^2} = g^{-1} + \frac{2}{L} \sum_{\vk} 
 \left[ d_{\vk}^2 + d_{\vk} u \frac{d n}{d\eta} \right] \,
 f'(\epsilon_{\vk} - \mu) \; .
\end{equation}
Exploiting the symmetry $n(\eta) = n(-\eta)$, which implies that 
odd derivatives of $n(\eta)$ vanish at $\eta=0$, one obtains
\begin{equation}
 a_2 = \left. \frac{d^2\omega}{d\eta^2} \right|_{\eta=0} = 
 g^{-1} + \frac{2}{L} \sum_{\vk} d_{\vk}^2 \,
 f'(\epsilon^0_{\vk} + u \, n(0) - \mu) =
 g^{-1} - N_2(\bar\mu,T)
\end{equation}
with $\bar\mu = \mu - u \, n(0)$.

Differentiating twice more and setting $\eta=0$, one obtains
\begin{equation}
 a_4 = \left. \frac{d^4\omega}{d\eta^4} \right|_{\eta=0} = 
 - N''_4(\bar\mu,T) + 3u \, N'_2(\bar\mu,T) 
 \left. \frac{d^2 n}{d\eta^2} \right|_{\eta=0} \; .
 \label{a4a}
\end{equation}
Applying two $\eta$-derivatives to $n(\eta)$ as given by \eq{selfn}
we get
\begin{equation}
 \left. \frac{d^2 n}{d\eta^2} \right|_{\eta=0} = N'_2(\bar\mu,T) - 
 u N_0(\bar\mu,T) \left. \frac{d^2 n}{d\eta^2} \right|_{\eta=0} \; .
\end{equation}
Solving for $n''(0)$ and inserting into \eq{a4a} we obtain \eq{a4}.


\vfill\eject


\end{document}